\documentclass{aastex631}

\shorttitle{Can Chameleon fields be the source of the  dark energy dipole and the CMB dipole?}
\shortauthors{Yarahmadi and Salehi}
\graphicspath{{./}{figures/}}

\begin{document}

\title{ \boldmath Can Chameleon fields be the source of the  dark energy dipole and the cmb dipole?}

\author[0000-0003-0954-4699]{Muhammad. Yarahmadi}
\affiliation{Lorestan University, 
Khorramabad, Iran}

\author{Amin Salehi}
\affiliation{Lorestan University, 
	Khorramabad, Iran}



\begin{abstract}

Recent research reveals that the Local Group is in motion towards $(l, b) = (276, 30)$ relative to the cosmic background radiation, manifesting a velocity of 600 $\frac{km}{s}$ a phenomenon recognized as the cosmic background radiation dipole or CMB dipole. Despite its well-documented nature, the precise cause of this peculiar motion remains elusive. High mass-density regions, such as galactic superclusters, stand out among the potential contributors to this cosmic flow. This paper employs chameleon fields to investigate anisotropies on both small and large scales. The data utilized in this study comprise Type Ia supernovae from the Pantheon catalog, totaling 1,048 supernovae within the redshift range of $0.015 < z < 2.3$. The analysis of bulk flow at various redshifts has yielded noteworthy discoveries. On a smaller scale (less than 150 Mpc), the movement direction of the Local Group coincides with that of the bulk flow. On a larger scale, the bulk flow direction corresponds to the direction of the dark energy dipole. This implies that the anisotropy at the local scale originates from the same source as the anisotropy observed on a larger scale.
\end{abstract}

\keywords{Bulk Flow --- Dark energy dipole--- CMB dipole --- Chameleon model}


\section{Introduction} \label{sec:intro}

Several studies have been conducted on cosmic background anisotropy by various researchers. The movement of local group galaxies relative to the cosmic background radiation, also known as cosmic background radiation dipole or CMB dipole, was found to be the best justification for this anisotropy by (\cite{Conklin}; \cite{Henry}; and \cite{Smoot}). Further, many studies were conducted to measure the exact value of the local cluster velocity and the cause of this movement. Initially, it was assumed that the Virgo supercluster, located at a distance of 17 megaparsecs from the local group galaxy, was responsible for this movement. However, Villumsen and Strauss found in 1987 (\cite{Villumsen}) that a high-density mass distribution over long distances was needed to justify the motion of a local cluster. Studies done by Shaya, Tammann (\cite{Shaya}; \cite{Tammann}), and Aaronson in 1986 (\cite{Aaronson}) illustrated that the difference in direction between the CMB dipole and the bulk flow was toward the center of Virgo in the direction of the Hydra-Centaurus superclusters, which are approximately located at a distance of 32 MPc. As a result, they reported that this supercluster attracts the local cluster.

In 1987, Dressier et al. (\cite{Dressier}) studied the motion of galaxies in the Hydra-Centaurus supercluster and showed that the supercluster was moving at a velocity of approximately $500 \, \mathrm{km \, s^{-1}}$ relative to the cosmic background radiation. As a result, this supercluster cannot be the sole reason for the local cluster motion, and a large mass distribution over a longer distance must be the source of this motion. Linden-Bell et al. (1988) (\cite{Lynden-Bell}) investigated the motion of 400 spiral galaxies that covered all angles of the sky and showed that these galaxies were moving toward a massive mass called the Great Attractor at a distance of $50-70 \, h^{-1} \mathrm{Mpc}$ and galactic coordinates $(l, b) = (325, -7.25)$.

In 1992, Bothun et al. (\cite{Bothun}) studied 48 spiral galaxies adjacent to the Great Attractor using the Tully-Fisher relation to demonstrate that these galaxies were moving toward the Great Attractor. After the first COBE satellite results were released, accurate velocity measurements of the local group galaxy were carried out by Kogut et al. \cite{Kogut} in 1993. Using the first data from the satellite, they estimated the velocity of the local group at $627 \, \mathrm{km \, s^{-1}}$. With the development of catalogs such as $(2MASS)$, Hoffman et al. (\cite{Hoffman}) in 2001, and Kocevski and Abiling in 2006 (\cite{Kocevski}), studies indicated that to determine the local group's dipole motion, the local group must be studied for distances farther away from the Shapley supercluster at $150 \, h^{-1} \mathrm{Mpc}$. Although studies conducted up to 2008 had a good convergence in the reconstruction of the motion of the local group galaxy, these results usually revealed a directional difference of 10 to 30 degrees. On the other hand, studies that sought to find the direction of motion of local group galaxies using the peculiar velocity of galaxies did not reach a definite conclusion to justify the motion of local group galaxies; they concluded, however, that galaxy clusters posed a collective and cohesive movement and that they have a particular direction to the cosmic background radiation called the bulk flow.

Until 2008, studies were conducted at depths of $150 \, h^{-1} \mathrm{Mpc}$, but in 2008 \cite{k}, Kashlinsky, Atrio-Barandela, Kocevski, and Ebeling analyzed three-year WMAP data and used the kinematics of Sunyaev–Zeldovich's data, and data covering up to $300 \, h^{-1} \mathrm{Mpc}$, and provided some evidence. The coordinated motion and coherent flow of galaxy clusters at velocities of $600-1000 \, \mathrm{km \, s^{-1}}$ were found in the direction of the galaxy coordinates, which were close to the constellations Centaurus and Vela. Researchers have speculated that this move may be a remnant of the impact of pre-inflation invisible regions of the universe. In 2009 and 2010, the results obtained by Kashlinsky et al. (\cite{k1};\cite{k2}) were corrected by Watkins, Feldman, and Hudson.

In 2013, data from the Planck satellite showed no evidence of bulk flow at this scale and claims that there is no evidence of gravitational effects beyond the visible universe. Using Planck and WMAP's data, they have seen evidence of bulk flow; thus, studies on finding the direction of motion of galaxies in the local group have posed a newer enigma called the bulk flow, for which a definitive answer has not been determined yet.

Most studies of bulk flow pursue three main goals:
\begin{enumerate}
	\item Calculate the direction and value of the bulk velocity.
	\item Determine the degree of convergence of the velocity of the bulk with the direction of motion of the galaxies in the local group.
	\item Investigate the degree of convergence for the motion of the bulk and the motion of the galaxies in the local group with superclusters such as Virgo, Hydra, Centaurus, Coma, and Shapley because it is thought that these superclusters, which are located at different distances, are the source of the bulk flow and the direction of local cluster motion.
\end{enumerate}

In this paper, in addition to the above, we determine the range of parameters of the Chameleon model using the bulk flow and the degree of convergence of the bulk flow with the dipole direction of dark energy.

\section{Theoretical model} \label{sec:style}

The bulk velocity vector can be visualized as the vector average of the full three-dimensional velocity field, computed over some chosen volume. If the velocity field can be sampled with arbitrary density, the bulk flow will be primarily sensitive to the distribution of mass outside or at the edges of the survey volume. It is for this reason that the bulk velocity is a sensitive probe of the mass clustering power on very large scales and potentially a discriminant between cosmological structure-formation models.

As mentioned, the bulk flow can be considered as a coherent motion of a large part of the universe. This motion includes the mass movement of galaxies, galaxy clusters, supernovae, and other cosmic objects relative to a reference frame, usually the cosmic background radiation. Different definitions and formulations of bulk velocity have been reported that may seem different in appearance but are fundamentally the same. Nusser et al. \cite{Nusser} have put forward a way to determine the velocity of the bulk through the following equation:

\begin{equation}\label{is}
	B(r)=\frac{3}{4\pi r^{3}}\int_{x<r}v(x)d^{3}x,
\end{equation}
\(v(x)\) is the peculiar velocity of objects within the volume of the sphere under study with radius \(r\). More often, different samples such as galaxy clusters and supernovae are studied. Although it is expected to measure the velocity of the bulk for all samples, they lead to almost the same answer; notwithstanding the more uniform the sample distribution, the less computational error, and different methods and catalogs for measuring the velocity of the bulk. The peculiar motion of supernovae causes changes in their luminosity distance and perturbation in their luminosity distance, which is expressed by the following formula: (\cite{Bon})

\begin{equation}\label{is}
	d_{L}(z,\upsilon_{DF},\theta)=d^{0}_{L}(z)+d^{dipole}_{L}(z,\upsilon_{DF},\theta).
\end{equation}

Where $d^{0}_{L}(z)$ is the luminosity distance of the supernova  and  obtained from the following relation:

\begin{equation}\label{is0}
	d^{0}_{L}(z)=c(1+z)\int^{z}_{0}\frac{dz^{\prime}}{H(z^{\prime})}.
\end{equation}.

In the above relation, $H(z)$, the Hubble parameter and $z$ redshift, $\upsilon_{DF}$ is the velocity of the bulk and $d^{(dipole)}_{L}(z,\upsilon_{DF},\theta)$ is the dipole domain of the velocity of bulk, which is obtained from the following equation:

\begin{equation}\label{is2}
	d^{(dipole)}_{L}(z,\upsilon_{DF},\theta)=\frac{\upsilon_{DF}(1+z)^{2}}{H(z)}\cos(\theta).
\end{equation}

Where \(\theta\) is the angle between the supernova line of sight and the dipole direction (bulk flow direction). It should be noted that the luminosity interval may be disturbed for other reasons, but other sources of disturbance are important in redshift greater than 1. (\cite{Hui}) Considering that in this article we have used 211 supernovae that are less than 0.1 in redshift, other sources of disturbance can be omitted.

It is important to note that although the bipolar fitting method (so-called dipole fit) seems to be a model-independent method, looking at the equation, it can be seen that this method is completely similar to the cosmological model through the Hubble parameter. Because the evolution of the Hubble parameter depends on the cosmological model and this parameter is one of the most basic cosmic parameters that is analyzed and fitted in different models both directly and through other parameters that are related to this parameter.

\section{Chameleon fields} \label{sec:floats}

Evidence for the accelerated expansion of the universe and time-dependent changes in the fine structure constant has led to the introduction of a scalar field with mass from order $H_{0}$, which is the Hubble flow constant value. If this field exists, it must be coupled with matter to satisfy the Equivalence principle. In 2003, Justin Khoury and Amanda Weltman (\cite{Khoury}) studied these fields and contended that the mass associated with these fields depends on the local density of matter in the universe. They also showed that the interaction range of these fields on Earth is about $1 \, \text{mm}$ and on a scale of the solar system is of the order of $10-10^4 \, \text{AU}$. All the limits and principles of general relativity have been included in these studies, predicting that Newton's constant measuring probes like SEE will obtain a different value from the order of one to its current value on the ground.

Considering the normalized action of Hilbert-Einstein as:

\begin{equation}\label{is0}
	S_{EH}=\int d^4 x\sqrt{-g}\frac{1}{16\pi G}R=\int d^4 x\sqrt{-g}\frac{M_{pl}^2}{2}R.
\end{equation}

In the presence of a scalar field $\phi$  and potential $V(\phi)$, another term as:

\begin{equation}\label{is0}
	S_\phi=-\int d^4 x\sqrt{-g}\left(\frac{1}{2}(\partial\phi)^2+V(\phi)\right),
\end{equation}

is added to the action. To introduce Chameleon fields, a set of fields $\psi_{m}^{(i)}$ by action:

\begin{equation}\label{is0}
	S_{m}=-\int d^4 x \mathcal{L}_{m}\left(\psi_{m}^{(i)},g_{\mu\nu}^{(i)}\right),
\end{equation}

which are coupled to scalar field $\phi$ , is added to them. In these relationships:

\begin{equation}
	g_{\mu\nu}^{(i)}\equiv e^{2\beta_i\phi/M_{pl}}g_{\mu\nu},
	\label{conformalrelation}
\end{equation}

$\beta_i$ is also a coupling constant for every small element of matter. So the total action is expressed as follows:

\begin{equation}
		S=\int d^4 x\sqrt{-g}\\
		\left(\frac{M_{pl}^2}{2}R-\frac{1}{2}\nabla_\mu\phi\nabla^\mu\phi-V(\phi)
		-\frac{1}{\sqrt{-g}}\mathcal{L}_{m}\left(\psi_{m}^{(i)},g_{\mu\nu}^{(i)}\right)\right),\\
\end{equation}

$\phi$, the equation of motion of the Chameleons is obtained as follows:

\begin{equation}
	\nabla^2\phi=V_{{eff},\phi}(\phi).
	\label{conciseeom}
\end{equation}

For a non-relativistic matter$(w_i\approx0)$, the effective potential is simplified by the following equation:

\begin{equation}
	V_{{eff}}(\phi)=V(\phi)+\sum_i\rho_i e^{\beta_i\phi/M_{pl}}.
\end{equation}

Since exponential potentials play an important role in describing the period of inflation, (\cite{Ferreira};\cite{Wetterich}) we examine the Chameleon field equations in the presence of potential $V = V_0 \exp\big (-\alpha\frac{\phi}{M_{pl}}\big)$.
Here $\alpha$ is a dimensionless constant that describes the potential slope. The value of $ \alpha$ and $\beta$ are plotted in Figure 7. These two parameters play a fundamental role in the dynamics of the cosmos. Another characteristic of these two parameters is that they determine the stability range of critical states in the chameleon model.\cite{sal} The Chameleon field equation in the presence of exponential potential is obtained as follows:

\begin{equation}
	\ddot{\phi}+3H\dot{\phi}=\frac{V_0\alpha}{M_{pl}} e^{\frac{-\alpha}{M_{pl}}\phi}-\frac{\beta}{M_{pl}}\rho_{m}e^{\frac{\beta}{M_{pl}}\phi}.
	\label{cosmiceom}
\end{equation}

Here $H$ is the Hubble parameter which is determined by the Friedmann constraint as follows:

\begin{equation}\label{Friedmann}
	3H^2M_{pl}^2=\frac{1}{2}\dot{\phi}^2+V_0e^{\frac{-\alpha}{M_{pl}}\phi}+\rho_{m}e^{\beta\phi/M_{pl}}
\end{equation}

\begin{equation}\label{Friedmann}
	(2\dot{H}+3H^2)M_{pl}^2=-\frac{1}{2}\dot{\phi}^2+V_0e^{\frac{-\alpha}{M_{pl}}\phi}-\gamma\rho_{m}e^{\beta\phi/M_{pl}}.
\end{equation}

The effective energy density in the Chameleon field is $\rho_{\text{eff}} = \frac{1}{2}\dot{\phi}^2 + V(\phi) + \rho_{m}e^{\beta\phi/M_{\text{pl}}}$, and the effective pressure is $P_{\text{eff}} = -\frac{1}{2}\dot{\phi}^2 + V(\phi)$. In previous studies, important parameters $(\alpha, \beta)$ in the Chameleon field have been investigated, and the range of these parameters has been fitted with observational tests (\cite{sal}; \cite{Burrage}). These two parameters play a crucial role in the dynamics of the universe. Another feature of these two parameters is that the stability range of critical states of the Chameleon model is determined by them. 
Many studies have investigated the chameleon parameters. The research article entitled 'Tests of Chameleon Gravity' by  (\cite{Burrage}) presents a compelling investigation of the Chameleon model. The study provides a comprehensive review of numerous experiments conducted to constrain the parameters of the model, with a particular focus on power law potentials and exponential couplings. This research asserts that the coupling value to matter should optimally be approximately unity.  The findings of this research may have significant implications for future research in this field. The application of screening mechanisms has yielded theories of gravity that conform to the predictions of general relativity within our solar system. These theories are nevertheless falsifiable through novel techniques such as the study of astrophysical phenomena in remote galaxies, or through targeted laboratory experiments. Furthermore, they may prove relevant for linear and non-linear cosmological scales in multiple cases. In our upcoming discussion, we will need to solve numerical calculations involving a nonlinear set of second-order differential equations (12-14). To simplify these calculations, we will convert this set of equations into first-order equations. This conversion is necessary for several reasons. Firstly, systems of first-order equations are much simpler to solve numerically, making computations more convenient and straightforward. Additionally, this transformation allows us to explore the behavior of the system in phase space, providing valuable insights into how the system changes over time. By reducing the equations to first-order, we only require a single initial condition for each equation, as opposed to multiple initial conditions needed for high-order equations. This simplifies the numerical solution process and makes it more efficient. Most importantly, describing the system's dynamics in terms of first-order equations enables us to comprehensively represent the system's behavior in phase space. This allows us to visualize the system's trajectory and assess its stability, providing a powerful tool for understanding its long-term evolution and equilibrium states.

\color{black}
Therefore, they can be examined in terms of numerical calculations. In this article, these variables are defined as follows:

\begin{equation}
	\Theta={\frac{\rho_{m}e^{\frac{\beta\phi}{M_{pl}}}}{3 H^{2}M_{pl}^{2}}} ,\ \	 \Upsilon=\frac{\dot{\phi}}{\sqrt{6} HM_{pl}},\ \  \zeta=\frac{V}{3H^{2}M_{pl}^{2}}.
\end{equation}
Considering equations (13) and (15), we can introduce Friedman's constraint as follows:

\begin{equation}\label{is00}
	\Theta+\Upsilon^2+\zeta=1.
\end{equation}\\

Therefore, the system equations are simplified as follows:

\begin{eqnarray}
	\frac{d\Theta}{dN}=\sqrt{6}\beta\Theta\Upsilon-2\Theta\frac{\dot{H}}{H^{2}},\label{y1}\\
	\frac{d\Upsilon}{dN}=-3\Upsilon+\frac{3\alpha}{\sqrt{6}}\zeta-\frac{3\beta}{\sqrt{6}}\Theta
	\Upsilon\frac{\dot{H}}{H^{2}},\label{x1}\\
	\frac{d\zeta}{dN}=-\sqrt{6}\alpha\Upsilon\zeta-2\zeta\frac{\dot{H}}{H^{2}}.\label{z1}
\end{eqnarray}

Which is $N\equiv\ln(a)$. \textquotedblleft Using constraint (16), the above equations are reduced to the following two equations:
\textquotedblright
Also, $\frac{\dot{H}}{H^{2}}$ is calculated as follows in terms of new variables:

\begin{eqnarray}
	\frac{\dot{H}}{H^{2}}=\frac{3}{2}\left[-1-\gamma\Theta-\Upsilon^2+\zeta\right].
\end{eqnarray}

The above parameter is one of the basic parameters through which both the equation of state is expressed and the calculation of the luminosity distance is made. The equation related to the luminosity distance is coupled with this equation with the system equations.  
Therefore, to find the luminosity distance and finally calculate the bulk flow, the following equations must be solved simultaneously. 

In order to incorporate the equation (3) with the dynamical system equations of (17-19), it can be rewriten in terms of the following differential equations
\begin{eqnarray}\label{dl2}
	\frac{dd^{0}_{L}}{dN}&=&-d^{0}_{L}-\frac{e^{-2N}}{H},  \\
	\frac{dH}{dN}&=&-H(\frac{\dot{H}}{H^{2}})\label{dl3}
\end{eqnarray}
Where, since $1+z\equiv\frac{1}{a}$, then $(1+z)\equiv e^{-N}$, $dz\equiv-e^{-N}dN$ and $dN\equiv Hdt$. The equations (\ref{dl2})- (\ref{dl3}) can be related to the equations (17-19) by $\frac{\dot{H}}{H^{2}}$ which have been obtained in terms on new variables. Hence in order to find $d^{0}_{L}$ and $H$ the set of equations (\ref{dl2})- (\ref{dl3}) must be coupled to (17-19). Hence the cosmological equations describing the dynamics of the anisometropic universe can be rewritten as the equations (17-22). 
These equations are a set of coupled first order differential equations (ODE) which can be solved numerically. 

The obtained $d^{0}_{L}$ in above equations  
is used to find $\mu_{th}(z)=5\log(d_{L})+42.38$.
we can compute the $\chi^2$-statistics for each case. Therefore, we proceed to define the following quantities:
\begin{equation}
	\chi^{2}_{\text{Pantheon}}= \sum_{i}^{N_{\text{Pantheon}} }\frac{    \left (  \mu(z_{i})_{\text{obs}}- \mu(z_{i})_{\text{th}} \right )^{2}    }{\sigma(z_{i})_{\text{obs,Pantheon}}^{2}},\\
\end{equation}
where $N$ is the number of data points, $\sigma_{i}$ is the uncertainty associated with each measurement.
\color{black}
\section{Dark Energy Dipole}
We employ the deviation in distance modulus from its optimal $\Lambda$CDM value as follows:
\begin{equation}
	\left( \frac{\Delta \mu (z)}{\bar \mu (z)} \right) \equiv \frac{\bar \mu (z) - \mu (z)}{\bar \mu (z)}
\end{equation}
where $\bar \mu$ denotes the best fit distance modulus within the framework of $\Lambda$CDM. The $1048$ Supernovae Type Ia (SnIa) data points from the Pantheon dataset are expressed in terms of distance moduli:
\begin{equation}
	\mu_{\text{obs}}(z_i) \equiv m_{\text{obs}}(z_i) - M \label{mug}
\end{equation}
Here, $m_{\text{obs}}$ represents the apparent magnitude of each SnIa, and $M$ is the assumed common absolute magnitude after proper calibration.

The best fit distance modulus, denoted as $\bar \mu(z)$, is determined by minimizing the following expression:
\begin{equation}
	\chi^2 (\Omega_m,\mu_0)= \sum_{i=1}^{1048} \frac{\left[\mu_{\text{obs}}(z_i) - \mu_{\text{th}}(z_i)\right]^2}{\sigma_{\mu \; i}^2}
	\label{chi2isotr}
\end{equation}
where $\sigma_{\mu \; i}^2$ represents the uncertainties in distance modulus, encompassing both observational and intrinsic random magnitude scatter. The theoretical distance modulus is defined as:
\begin{equation}
	\mu_{\text{th}}(z_i) \equiv m_{\text{th}}(z_i) - M = 5 \log_{10} (D_L (z)) +\mu_0 \label{mth}
\end{equation}
where $\mu_0$ is a constant related to the Hubble parameter $H_0 \equiv 100\;h$ km/(sec $\cdot$ Mpc) \cite{Sanchez:2009ka} by:
\begin{equation}
	\mu_0 = 42.38 - 5 \log_{10}h \label{mu0}
\end{equation}
and
\begin{equation}
	D_L (z) = (1+z) \int_0^z dz'\frac{H_0}{H(z';\Omega_m)} \label{dlth1}
\end{equation}
is the Hubble free luminosity distance. Minimizing $\chi^2 (\Omega_m,\mu_0)$ using the Pantheon dataset yields the best fit parameter values $\Omega_m=0.283\pm 0.044$ and ${{{\bar \mu }_0}}=42.36\pm0.04$, completely specifying $\bar \mu(z_i)$. Consequently,
\begin{equation}
	\left( \frac{\Delta \mu (z_i)}{\bar \mu (z_i)} \right)_{\text{obs}} \equiv \frac{\bar \mu(z_i) - \mu(z_i)}{\bar \mu(z_i)} \label{dmmudef}
\end{equation}
for all Pantheon datapoints.

The direction of the dark energy dipole is determined to be ($b=-15.1^\circ \pm 11.5^\circ$, $l=309.4^\circ \pm 18.0^\circ$).

\section{CMB Dipole}

The Cosmic Microwave Background (CMB) dipole is the most prominent expression of anisotropy. The amplitude of the CMB dipole registers at approximately $3.3621\pm0.0010$ mK (\cite{Henry}). Presuming the universe's homogeneity and isotropy, an observer anticipates a blackbody spectrum with temperature $T$ across the celestial expanse. The dipole spectrum has substantiated its identity as the derivative of a blackbody spectrum.

Noteworthy is the frame-dependent nature of the CMB dipole. It can be construed as the Earth's peculiar motion towards the CMB. The amplitude undergoes temporal variation owing to the Earth's orbital trajectory around the solar system's barycenter. This temporal evolution prompts the incorporation of a time-dependent term into the dipole expression, exhibiting a modulating period of one year, a characteristic corroborated by observations conducted with COBE FIRAS. Importantly, the dipole moment bears no imprint of primordial information.

Analysis of CMB data reveals that the Sun exhibits an apparent motion at $368\pm2$ kilometers per second relative to the CMB reference frame, also designated as the CMB rest frame, wherein no motion is perceived through the CMB. Furthermore, the Local Group, a galaxy cluster encompassing our Milky Way, appears to traverse at $627\pm22$ kilometers per second in the direction of galactic longitude $ l = 276 \pm 3,b = 30 \pm 3 $ (\cite{Kogut}). This motion engenders anisotropy in the data, with the CMB manifesting a slightly elevated temperature in the direction of movement compared to the opposite direction. The prevailing interpretation attributes this temperature disparity to a straightforward velocity redshift and blueshift resulting from motion relative to the CMB. However, alternative cosmological models offer explanations for a fraction of the observed dipole temperature distribution in the CMB.

\subsection{Numerical analysis and simulation} \label{subsec:tables}

In this article, we use the Pantheon catalog, which contains information on redshift, distance modulus, measurement error, and galactic coordinate of 1048 supernovae in redshift less than 2.3. If $\hat{n}_{i}$ is a single vector in the direction of the supernova $i$ of view, then:

\begin{equation}
	\hat{n_{i}}= \cos(l_{i})\sin(b_{i})\hat{i}+\sin (l_{i}) \sin( b_{i})\hat{j}+\cos (b_{i})\hat{k}.
\end{equation}

Where $(l_{i},b_{i})$ is the galactic coordinates of the supernova $i$. Similarly, if $\hat{p}$ is a unit vector in the dipole direction, then,

\begin{equation}
	\hat{p}= \cos(l)\sin (b)\hat{i}+\sin (l) \sin( b)\hat{j}+\cos(b)\hat{k}.
\end{equation}

Which $(l,b)$ is the galactic coordinates of the bulk flow. Thus

\begin{equation}
		cos\theta_{i}=(\hat{n_{i}}.\hat{p})=\cos(l)\sin (b)\cos(l_{i})\sin(b_{i})\\
		+\sin (l) \sin( b)\sin (l_{i}) \sin( b_{i})+\cos(b)\cos (b_{i}).
\end{equation}

We use the $\chi^{2}$ method to calculate the direction and value of the bulk flow velocity

\begin{equation}\label{is}
	\chi^{2}=\sum_{i}\frac{|\mu_{i}-5\log_{10}((d^{0}_{L}(z_{i})-d^{dipole}_{L}(z,\upsilon_{DF},\theta_{i})/10 pc|^{2}}{\sigma^{2}_{i}}.
\end{equation}

The distance modulus $\mu_{i}$ is obtained from the following formula:

\begin{eqnarray}\label{distancem}
	\mu _{i}=5\log_{10} d_{L}(z)+42.384-5\log_{10} h_{0}.
\end{eqnarray}

\begin{itemize}
	\item $d^{0}_{L}(z_{i})$: Denotes the luminosity distance at redshift $z_{i}$ in a reference cosmology.
	\item $d^{dipole}_{L}(z, \upsilon_{DF}, \theta_{i})$: Refers to the luminosity distance, considering the dipole component of the bulk flow. It involves parameters such as the dipole amplitude ($\upsilon_{DF}$) and the angle ($\theta_{i}$).
	\item $\sigma_{i}$: Represents the uncertainty or standard deviation associated with the observation at redshift $z_{i}$.
\end{itemize}

The chi-square ($\chi^2$) is computed as the sum of the squared differences between the observed magnitude and the expected magnitude (based on luminosity distances and the dipole component), divided by the squared uncertainty for each data point ($\sigma_{i}^2$).

To derive the direction and value of the bulk flow velocity, we likely employ a fitting or optimization procedure to minimize the overall chi-square statistic with respect to relevant parameters, such as the dipole amplitude ($\upsilon_{DF}$) and the angle ($\theta_{i}$). The minimum value of $\chi^2$ corresponds to the best-fit values for these parameters, providing information about the bulk flow velocity in terms of its direction and magnitude.

Hubble constant $h_{0}$ in unit $100kms^{-1} Mpc^{-1}$ for three different redshift, we do the numerical solution of the equations.\\
1. Redshift interval less than 0.035 is equivalent to an approximate distance of 140$h^{-1}Mpc$\\
2. Redshift interval less than 0.06 is equivalent to an approximate distance of 250$h^{-1}Mpc$\\
3. Redshift interval less than 0.1 is equivalent to an approximate distance of 422$h^{-1}Mpc$

\section{Redshift interval less than 0.035}

From 1,048 supernovae, 124 supernovae are in redshift less than 0.035. The parameters to be fitted are $(l,b,\upsilon_{DF})$ and the parameters of the Chameleon model $(\alpha,\beta)$. In Figure 1, the probability distribution function of the bulk flow is plotted in galactic coordinates. The simulation results with data greater than $2\times10^5$ indicate that the bulk flow is moving in the direction $(l,b)=(286^{+14}_{-12},16^{+12}_{-8})$ at a velocity of $273kms^{-1}$. In this redshift, the Hydra-Centaurus supercluster exists. The bulk flow is approximately in the direction of this supercluster. This result is in good agreement with the results of Kashlinsky (\cite{k};\cite{k2};\cite{Yarahmadi}) studies who obtained the value $(l,b)=(287^{+9}_{-9},8^{+6}_{-6})$ for redshift less than 0.03 in the region $(1-\sigma)$. Feindt et al.(\cite{fien}) also used 117 supernovae in the redshift less than 0.1 in their research work and obtained the bulk flow velocity in the $(l,b)=(290^{+20}_{-20},15^{+18}_{-18})$ at a velocity of $292^{+96}_{-96}kms^{-1}$ region $0.015<z<0.035$, which is in good agreement with the result of our calculations in this paper in the $(1-\sigma)$ region. The shape of the $(1-\sigma)$ region corresponding to the direction of the bulk flow on a figure is shown in galactic coordinates. Moreover, the findings of some previous studies and the coordinates of important superclusters located in this redshift region are shown in Figure 2 (top panel). Another noteworthy point in Figure 2(low panel) is that the direction of motion of the local group galaxy is very close to the obtained bulk flow in this region.

Using the bulk flow model, the parameters of the Chameleon model $(\alpha,\beta)$ are fitted. 
\begin{figure}[tbp]
	\centering 
	\includegraphics[width=1\textwidth,height=.35\textheight]{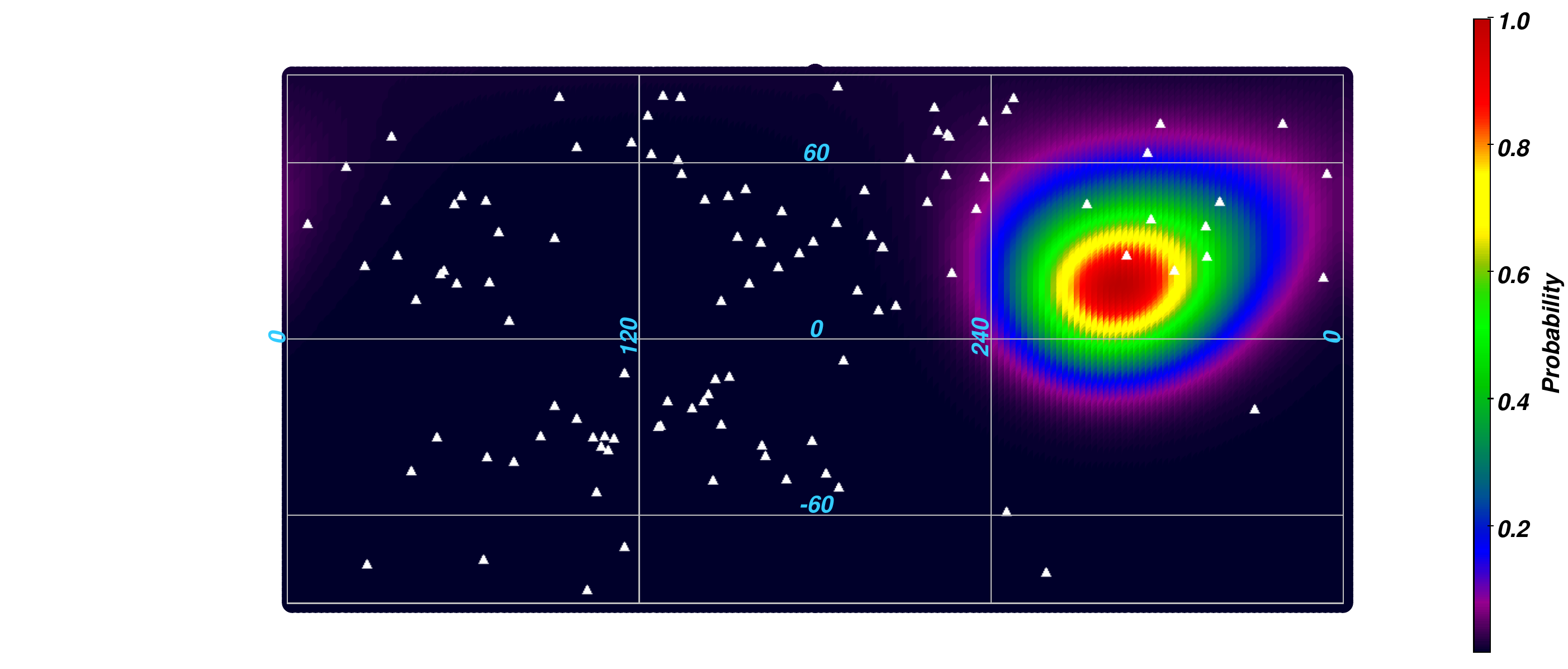}
	
	\caption{\label{fig:i} Probability of bulk flow direction in galactic longitude $l$ and galactic
		latitude $b$ for $0.015<z<0.035$. The most probable direction pointing towards $(l,b)=(286^{+14}_{-12},16^{+12}_{-8})$. The white triangles denotes the 124 Supernovas . The bulk flow direction in  this redshift is in direction of CMB dipole at $(l,b)=(276,30)$.
	}
\end{figure}
\begin{figure}[tbp]
	\centering 
	\includegraphics[width=.8\textwidth,height=.8\textheight]{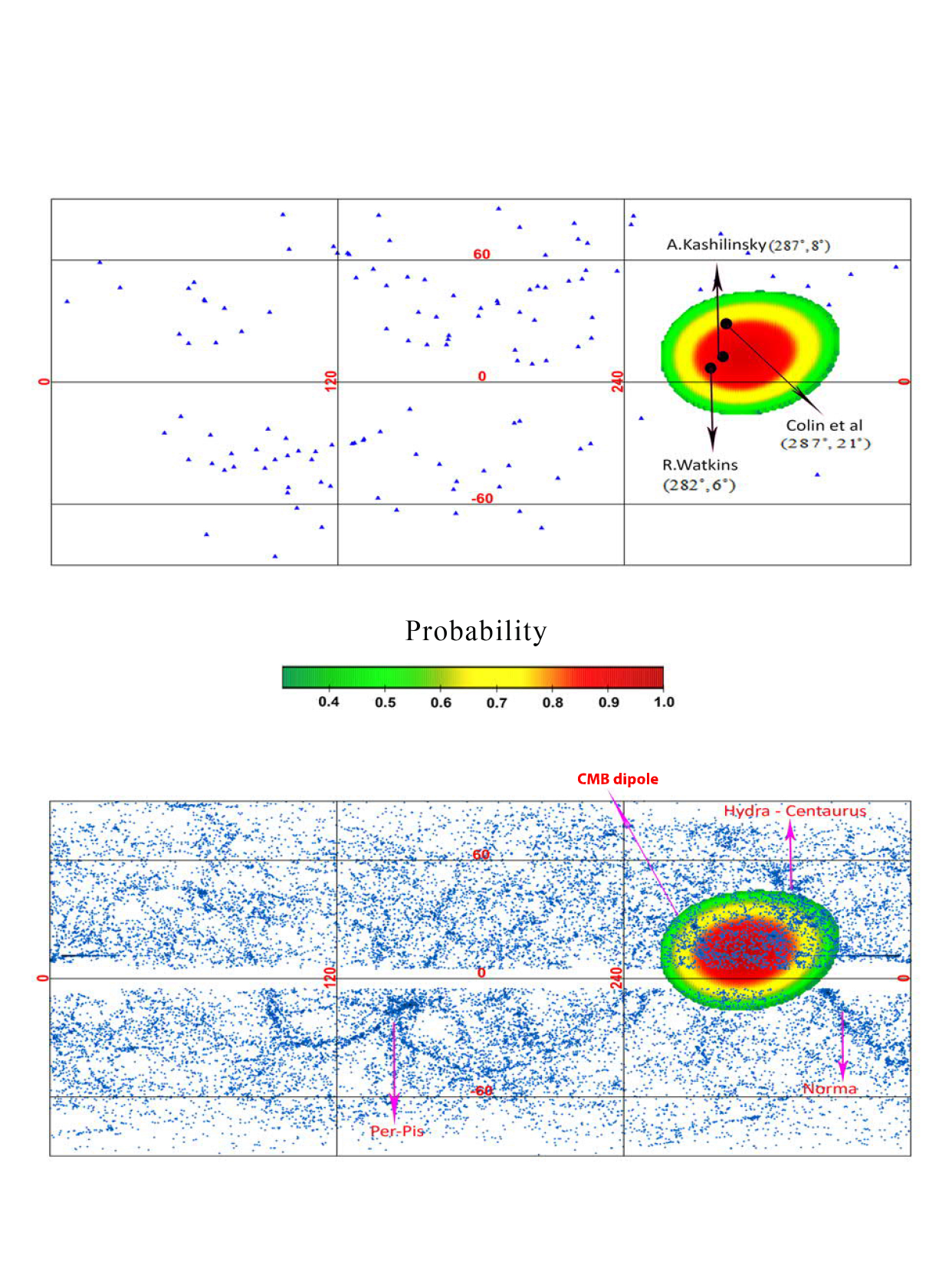}
	
	\caption{\label{fig:i}Top panel: Probability distribution function related to bulk flow direction, plotted in galactic coordinates and shown the results of some previous studies , in the redshift  less than 0.035. Blue triangles denotes supernovas . 
		Low panel: area $(1-\sigma)$ is plotted for the bulk flow direction plotted  in galactic coordinates for redshift less than 0.035. Also, Red arrows indicate that the positions of the superclusters. Blue circles denotes  distribution of galaxies in this redshift.  }
\end{figure}

\section{Redshift interval less than 0.06}

In the redshift $0.015<z<0.06$, there are 174 supernovae. In figure 2(top panel), the probability distribution function for the bulk flow  in galactic coordinates is plotted. The simulation results with more than $2\times 10^5$ data show that the bulk flow in this area is moving in the direction of $(l,b)=(300^{+15}_{-18},3^{+10}_{-11})$, which is close to the direction of the constellations Centaurus and Hydra. The Shapley supercluster is also located in the redshift and it is thought that all superclusters are moving towards this supercluster. This  is consistent with the results of studies (\cite{Lavaux}) ($(l,b)=(295^{+18}_{-18},14^{+18}_{-18})$)‌, \cite{Kocevski} ($(l,b)=(306.44,29.71)$) , (\cite{Feldman}) ($(l,b)=(282^{+11}_{-11},6^{+6}_{-6})$)  (\cite{De}) ($(l,b)=(290^{+39}_{-31},20^{+30}_{-30})$). Figure2  describes the region $(1-\sigma)$ corresponding to the direction of the bulk flow in the galactic coordinates. Another noteworthy point in figure 2(low panel) is that the direction of motion of the local group galaxy is at the edge of the region's confidence level $(1-\sigma)$ in the direction of the bulk flow, which is indicative of the fact that the reconstruction of the direction of motion of the local cluster by Bulk flow for distances less than $150h^{-1}Mpc$ leads to better results.
\begin{figure}[tbp]
	\centering 
	\includegraphics[width=1\textwidth,height=.3\textheight]{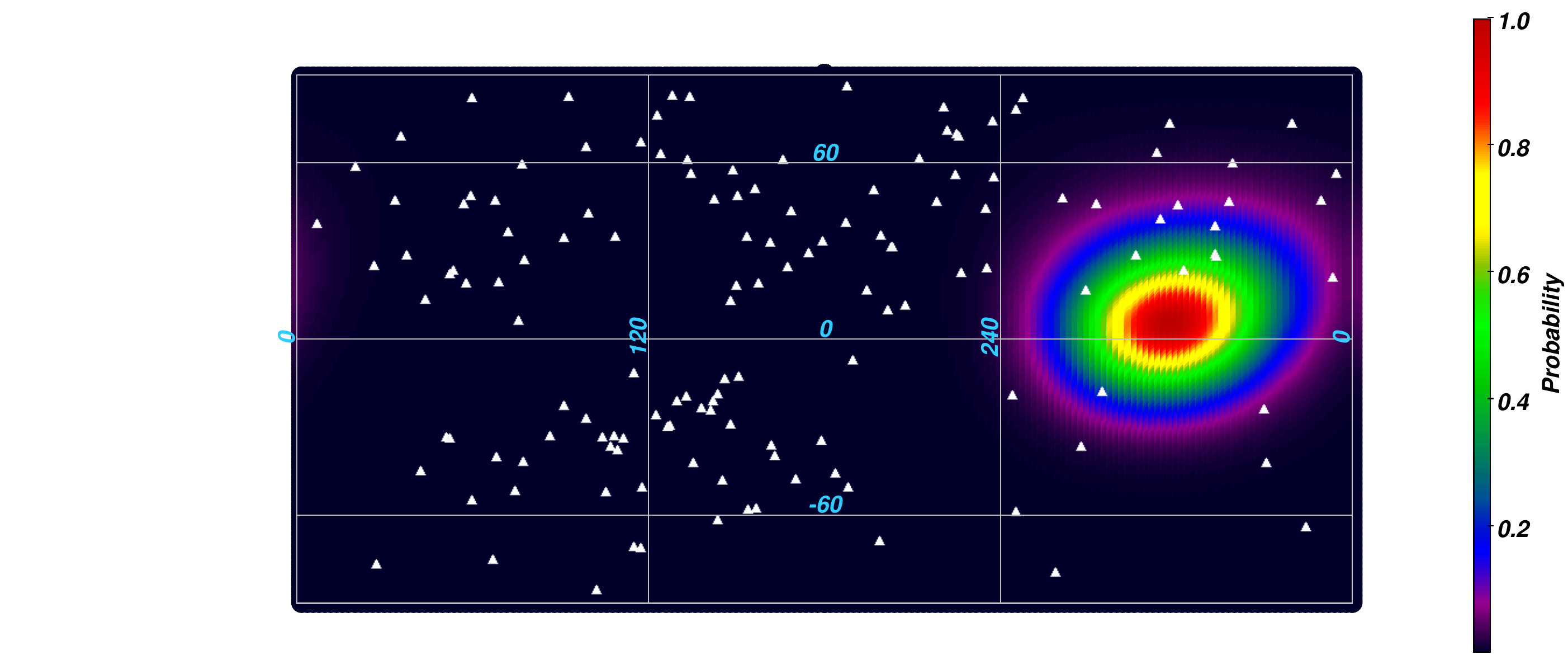}
	\caption{\label{fig:i}Probability of bulk flow direction in galactic longitude $l$ and galactic
		latitude $b$ for $0.015<z<0.06$. The most probable direction pointing towards $(l,b)=(300^{+15}_{-18},3^{+10}_{-11})$. The white triangle denotes the Supernovas. The white triangles denotes the 174 Supernovas . The bulk flow direction in  this redshift is in direction of the Shapely super cluster at $(l,b)=(311,32)$.}
\end{figure} 
\begin{figure}[tbp]
	\centering 
	\includegraphics[width=.8\textwidth,height=.8\textheight]{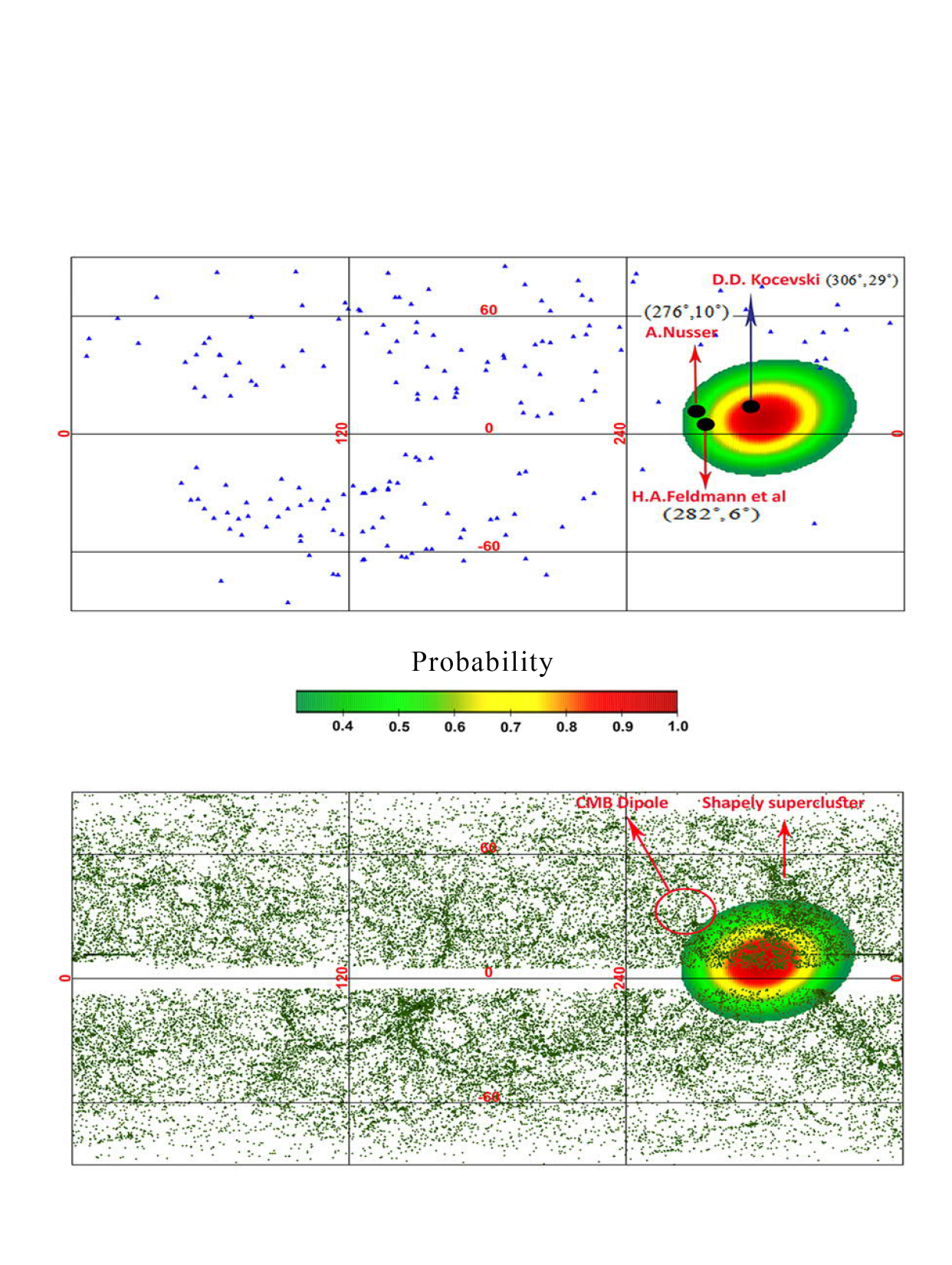}
	\caption{\label{fig:i}Top panel: Probability distribution function related to bulk flow direction, written on the rectangle in galactic coordinates and shown the results of some previous studies , in the redshift  less than 0.06. Blue triangles denote as supernovas. 
		low panel: area $(1-\sigma)$ is plotted for the bulk flow direction plotted in galactic coordinates for redshift less than 0.06. Also,  Red arrows indicate the indicators that display the positions of the superclusters. Green circles denotes  distribution of galaxies in this redshift.}
\end{figure} 

\section{Redshift interval less than 0.1} \label{subsubsec:autonumber}

In the redshift  $0.015<z<0.1$, there are 211 supernovae. In this redshift, the result  obtained for the bulk flow direction is $(l,b)=(302^{+10}_{-12},5^{+8}_{-9})$, which is consistent with the results of the (\cite{Watkins}) ($(l,b)=(283^{+14}_{-18},12^{+14}_{-14})$)‌. Figures (3) demonstrate the results of this simulation. As shown in Figure 3(low panel), the direction of movement of the local cluster (CMB dipole) is outside the region $(1-\sigma)$ of the bulk flow, which indicates that the farther we go from $150h^{-1}Mpc$, the difference between the direction of movement of the local cluster and the direction of the bulk flow increases. The results of many previous studies for different areas are given in Table (1).

\begin{figure}[tbp]
	\centering 
	\includegraphics[width=.8\textwidth,height=.6\textheight]{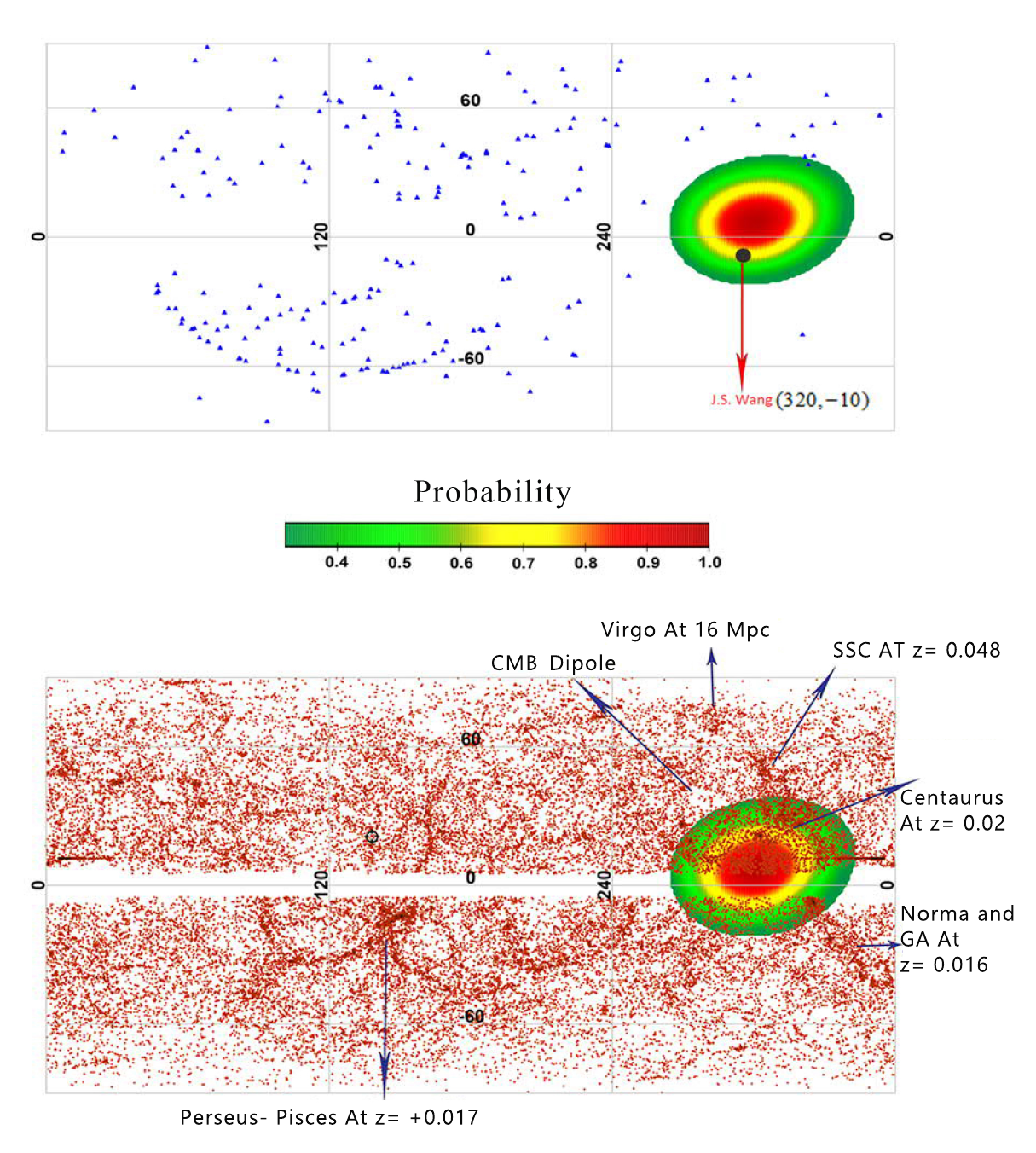}
	\caption{\label{fig:i}Top panel: Probability distribution function related to bulk flow direction, plotted in galactic coordinates and shown the results of some previous studies , in the redshift  less than 0.1. Blue triangles denote as supernovas. 
		Low panel: area $(1-\sigma)$ is plotted for the bulk flow direction plotted in galactic coordinates for redshift less than 0.1. Also, Black arrows indicate  that  the positions of the  all important superclusters in $0.015<z<0.1$. Orange circles denotes  distribution of galaxies in this redshift. }
\end{figure}

The other parameters that are fitted are the parameters of the Chameleon model $ (\alpha, \beta) $, which is determined using the bulk flow model. In Figure 4, the probability distributions for these parameters are plotted for 60,000 data. As it is known, the range of these parameters is of the order of one, which some studies have predicted the same value.
\begin{figure}[tbp]
	\centering 
	\includegraphics[width=0.4\textwidth,height=0.3\textheight]{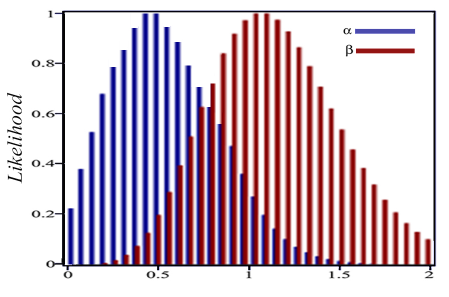}\\
	\caption{\label{fig:i}Probability distribution function related to Chameleon parameters $ (\alpha, \beta) $ fitted to supernovae type Ia in the Bulk flow model.}
\end{figure}
\begin{table}
	\scriptsize 
	\begin{center}
		\caption{List of studies on bulk flow}
		\label{table1}
		\begin{tabular}{llclccccl}
			\hline
			\hline 
			\rule{0pt}{8mm}
			$Ref$  & $velocity$ \ & $redshift$ \ & $l^{o}$ \ & $b^{o}$ \ & $distance$
			\\
			\rule{0pt}{8mm}
			$$  & $kms^{-1}$ \ & $$ \ & $degree$ \ & $degree$ \ & $h^{-1}Mpc$\\
			\hline 
			\rule{0pt}{8mm}
			\cite{k2} &$1000$ &$z\leq0.03$ & $287\pm9$ & $8\pm6$& $127$\\
			\hline 
			\rule{0pt}{8mm}%
			\cite{Watkins} &$ 407\pm81$ &$z\leq0.2$ & $283\pm14$ & $12\pm14$& $857$ \\
			\hline 
			\rule{0pt}{8mm}%
			\cite{Kocevski}  & $507$ & $0.035\leq z\leq0.055$ & $306.44$ & $29.71$& $127-220$ \\
			\hline 
			\rule{0pt}{8mm}
			\cite{Turnbull} &$249\pm 276$ & $z\leq0.2$ & $319\pm18$ & $7\pm14$&  $857$ \\
			\hline 
			\rule{0pt}{8mm}
			\cite{Colin}  & $250^{+190}_{-160}$ & $0.045<z<0.06$ & $287$ & $21$ & $168-249$ \\
			\hline 
			\rule{0pt}{8mm}
			\cite{Watkins} & $416\pm 78$ & $z=0.0167$ & $282$ & $60$& $45$ \\
			\hline 
			\rule{0pt}{8mm}
			\cite{Lavaux}  & $473\pm 128$ & $0.035<z<0.055$ & $220$ & $25$& $127-220$ \\
			\hline 
			\rule{0pt}{8mm}
			\cite{Nusser} & $257\pm 44$ & $0.035<z<0.055$ & $276\pm6$ & $10\pm6$& $127-220$ \\
			\hline 
			\rule{0pt}{8mm}
			\cite{Feldman} & $416\pm 78$ & $0.015<z<0.06$ & $282\pm11$ & $6\pm6$& $40-249$ \\
			\hline 
			\hline 
		\end{tabular}
	\end{center}	
\end{table}

\section{The relation between Chameleon model with cmb dipole and dark energy dipole}
In this section, we investigate the relation between Chameleon model with the dark energy and the CMB dipoles. Cosmic background radiation is the oldest light emitted from the childhood of the universe, and the photons of the last scattering level have been traveling ever since. By studying and analyzing cosmic background radiation data, scientists noticed temperature anisotropies in this radiation. Important information such as the distribution of matter and energy in the Universe can be obtained from these anisotropies . The CMB dipole is one of these anisotropies, and  one of the causes of which may be the movement of the Earth in space relative to the cosmic background radiation. By using this dipole, it is possible to calculate the peculiar velocity of local group galaxies and also put constraints on the large-scale structure and cosmic evolution.
Another anisotropy that can be observed from the cosmic background radiation is a dipole, which reveals that the energy in the Universe is distributed heterogeneously and there may be various reasons for it. The name of this anisotropy is dark energy dipole. In the following, we investigate the relation between chameleon field and dark energy dipole. The chameleon field changes its value depending on the mass of a region in space. In regions with a high mass of matter its mass is high, and in regions with a low mass of matter its mass changes and becomes lower. Thus, it can be concluded that the distribution of matter in the universe is heterogeneous. The chameleon field can explain the accelerated expansion of the universe with modifications in gravity. The chameleon particles can interact with matter. Thus, considering that matter is not uniformly distributed in the Universe, the chameleon field becomes heterogeneous. It seems that we can find the possibility of a relationship between the chameleon field and the dark energy dipole. If matter is non-uniformly distributed in the universe, this may lead to a heterogeneous distribution of dark energy. 
The last statement explains a  connection between the non-uniform distribution of cosmic matter and the resulting variation in dark energy distribution, influencing the overall structure of the universe. Implicit in this idea is the acknowledgment that the universe has a diverse spatial arrangement of matter, giving rise to various cosmic structures like galaxies, clusters, and cosmic voids. In the prevailing cosmological framework, dark energy plays a crucial role as the supposed driver of the observed accelerated expansion of the universe. The interaction between matter and dark energy unfolds as a complex cosmic interplay. Regions with higher matter density experience gravitational dominance, leading to a slowdown in cosmic expansion. In regions where the density of matter is relatively low, the effects of dark energy become more prominent. This causes an accelerated expansion of the universe due to the repulsive nature of dark energy. This dynamic equilibrium results in a diverse distribution of cosmic regions, each governed by a unique balance between matter and dark energy. The non-uniform distribution of matter, shaped by the intricate gravitational dynamics inherent in the cosmic structure, emerges as a fundamental factor in delineating the local manifestations of dark energy.  This nuanced relationship is crucial not only for understanding the observed large-scale structures in the universe but also for unraveling some intrinsic properties of dark energy. The elusiveness of the Chameleon field to detection arises from its unique properties and the mechanism it employs to remain hidden in certain environments. Its distinctive feature lies in its ability to dynamically adjust its mass in response to the surrounding matter density, known as the "chameleon mechanism."  In high-density environments, such as those found in laboratory settings or within massive celestial bodies like stars or planets, the Chameleon field acquires a substantial mass. This increased mass results in a shorter range of interaction, effectively suppressing its effects over short distances. As a consequence, experiments conducted in high-density environments are less likely to detect the Chameleon field. The Chameleon model incorporates a screening mechanism that shields the effects of the Chameleon field in regions of high matter density. This mechanism is crucial for the model to avoid conflict with precision tests of gravity conducted in environments like the Solar System. The Chameleon field effectively "screens" itself from detection in these regions, allowing it to remain consistent with established gravitational theories. On cosmological scales, where matter density is significantly lower, the Chameleon field exhibits a lighter mass. In such environments, its range of influence is extended, and its effects on cosmic expansion become more pronounced. However, these effects can be subtle and challenging to distinguish from other cosmological factors. Detecting the Chameleon field requires sophisticated experiments and precise measurements, with observational challenges arising from the need to differentiate the subtle effects of the Chameleon field from other astrophysical and cosmological phenomena. Experimental setups must be tailored to account for the field's elusive nature. There are experimental searches that constrain the chameleon coupling and other parameters by, for instance, using vacuum chambers, atom interferometry, and other techniques. Hence, in principle, one could detect the chameleon field, even on Earth (i.e. a high-density environment) by artificially creating a low-density environment (i.e. a vacuum chamber). Of course, such experiments can only gauge a certain part of the chameleon parameter space. Nonetheless, in principle, one could detect the chameleon experimentally, if it had certain coupling and energy scale parameter values that the currently existing experiments are sensitive to.

\color{black}
There are several ways to prove or disprove the  relation between distribution of matter and heterogeneous of dark energy  . One of these ways is to study the power spectrum of the cosmic background radiation.  Another possibility is to study the large-scale structure, and bulk flow can be used for this purpose.
 In this paper, we investigate the bulk flow to prove or refuse the relation between dark energy dipole and chameleon field. For this purpose, we use the redshift tomography method.

\section{Redshift Tomography}
Redshift tomography stands as an advanced method in cosmology, delving into the intricacies of the distribution of galaxies across distinct redshift ranges. Redshift, a consequence of the Doppler effect, serves as a measure of the universe's expansion, playing a pivotal role in unraveling the grand-scale structure and evolutionary trajectory of our cosmos.  At its core, redshift tomography aims to scrutinize the three-dimensional arrangement of matter in the universe across different cosmic epochs. This involves partitioning observed astronomical objects into distinct redshift intervals, providing a nuanced lens through which researchers can explore the transformations in cosmic structures over time.  In the realm of galaxy surveys, redshift tomography empowers the reconstruction of the density field of galaxies within various redshift bins. This not only facilitates an in-depth exploration of the development and expansion of substantial cosmic structures like galaxy clusters and filaments but also enables an examination of cosmic acceleration and the enigmatic properties of dark energy.  The potency of redshift tomography becomes even more apparent when intertwined with additional observational tools, such as cosmic microwave background (CMB) readings and gravitational lensing.  In redshift tomography, we split the data in different redshift shells, so that none of the shells are related to each other. We focus on regions that include objects with high matter density. We work in   $ 0.025<z<0.45 $, $ 0.045<z<0.55 $,$0.4<z<0.6 $, $0.1<z<1.4 $ redshift shells for calculating the bulk flow direction. As we mentioned above, the density of the chameleon field in regions with high matter density increase and cause non-uniform gravity which can create the preferred direction in the universe. 	This section focuses on examining the direction and velocity of the bulk flow across different redshift ranges. Several significant galactic superclusters, such as Perseus-Pisces (PP), Coma, Norma, and Virgo, are found within the redshift region of $0.015 < z < 0.035$. We are particularly interested in the redshift range of $0.016 < z < 0.03$, within which a giant supercluster called the 'Perseus-Pisces' (PP) supercluster is situated. The radial extent of this supercluster is estimated to be roughly 32 Mega parsec. Figure8, shows the direction of bulk flow in in this redshift.  In Figures (9), the direction of the PP supercluster in galactic coordinates is presented alongside the direction of the bulk flow obtained from the chameleon model. Our numerical simulations reveal that the bulk flow exhibits a direction characterized by galactic longitude ($l$) and galactic latitude ($b$) values of $(l, b) = (112^\circ \pm 12^\circ, -15^\circ \pm 13^\circ)$. As depicted in this figure, there is a good agreement between the direction of the bulk flow and the position of the PP supercluster.
This analysis sheds light on the correlation between the observed bulk flow and the spatial arrangement of the Perseus-Pisces supercluster, providing valuable insights into the dynamics and interactions within this cosmic structure.

\begin{figure}[tbp]
	\centering 
	\includegraphics[width=1.2\textwidth,height=.35\textheight]{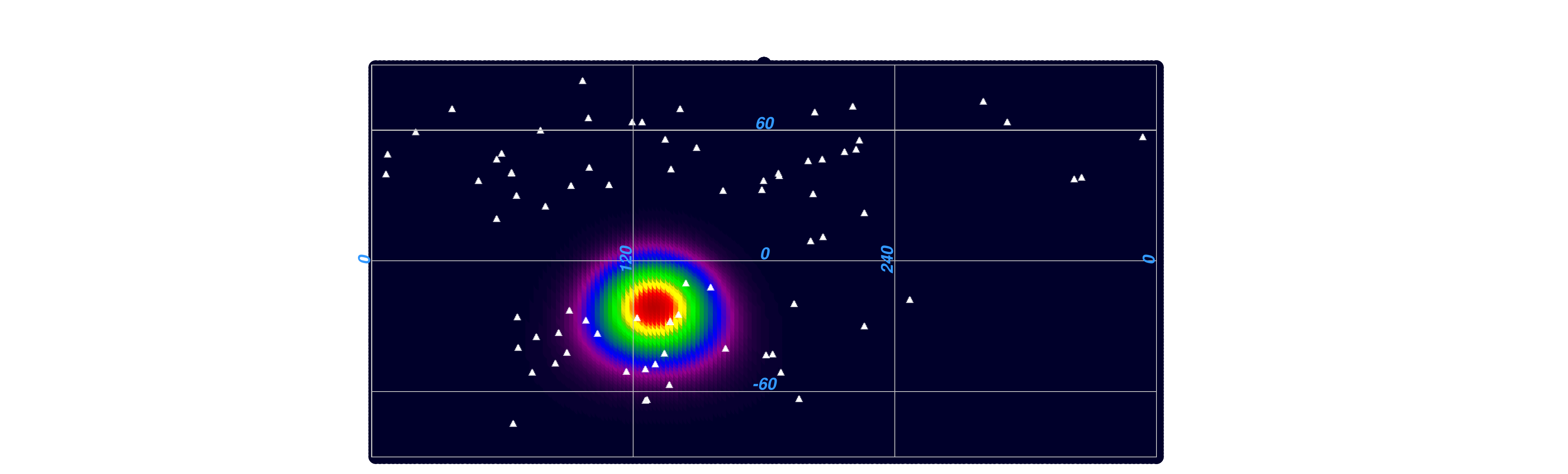}
	
	\caption{\label{fig:i} Probability of bulk flow direction in galactic longitude $l$ and galactic
		latitude $b$ for $0.016<z<0.03$. The most probable direction pointing towards $(l,b)=(112^{+14}_{-12},-16^{+12}_{-8})$. The white triangles denotes the Supernovas in this redshift range.
	}
\end{figure}
\begin{figure}
	\centering 
	\includegraphics[width=0.7\textwidth,height=.35\textheight]{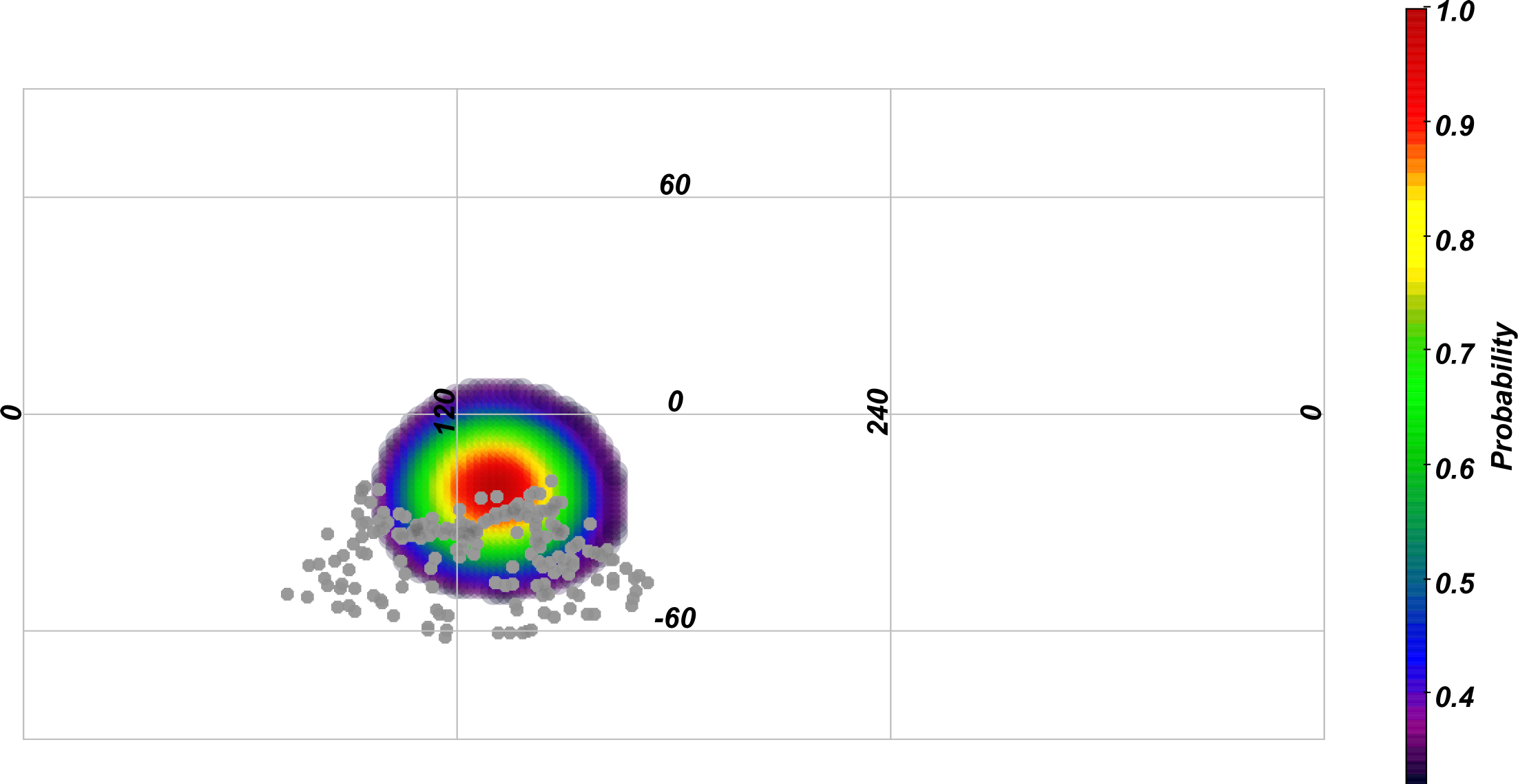}
	
	\caption{\label{fig:i} The probability distribution of bulk flow direction within the galactic coordinates of longitude ($l$) and latitude ($b$) is depicted for the redshift range $0.016 < z < 0.03$. The figure highlights the most probable direction, centered around $(l, b) = (112^{+14}_{-12}, -16^{+12}_{-8})$. The gray circle corresponds to the spatial extent of the Perseus-Pisces supercluster, providing contextual information for the observed bulk flow patterns. The PP position is in the direction of bulk flow in this redshift.
	}
\end{figure}

Shapley supercluster are in the redshift $0.45<z<0.055$. Figure 10 demonstrate the direction of bulk flow in this redshift. The Shapley Supercluster, also known as the Shapley Concentration, is one of the largest known concentrations of galaxies in the observable universe.  Galaxies within the Shapley Supercluster are not static; rather, they are in constant motion due to gravitational influences. As we can see in Fig 11, shapley super cluster position is in the (1-$\sigma$) with direction of bulk flow.

\begin{figure}[tbp]
	\centering 
	\includegraphics[width=0.7\textwidth,height=.35\textheight]{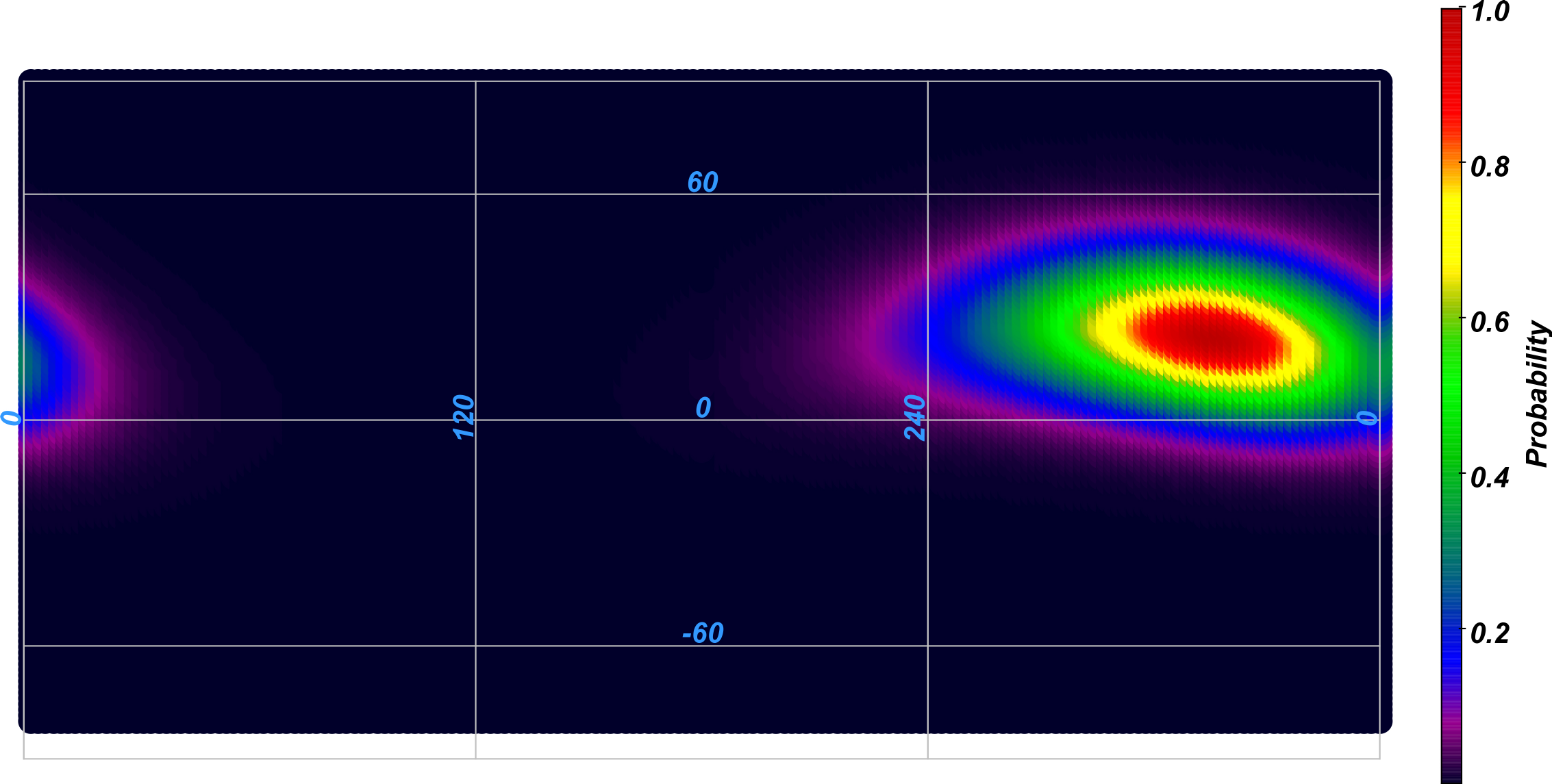}
	
	\caption{\label{fig:i} Probability of bulk flow direction in galactic longitude $l$ and galactic
		latitude $b$ for $0.045<z<0.055$. The most probable direction pointing towards $(l,b)=(312^{+14}_{-18},26^{+12}_{-8})$. 
	}
\end{figure}

\begin{figure}[tbp]
	\centering 
	\includegraphics[width=0.7\textwidth,height=.35\textheight]{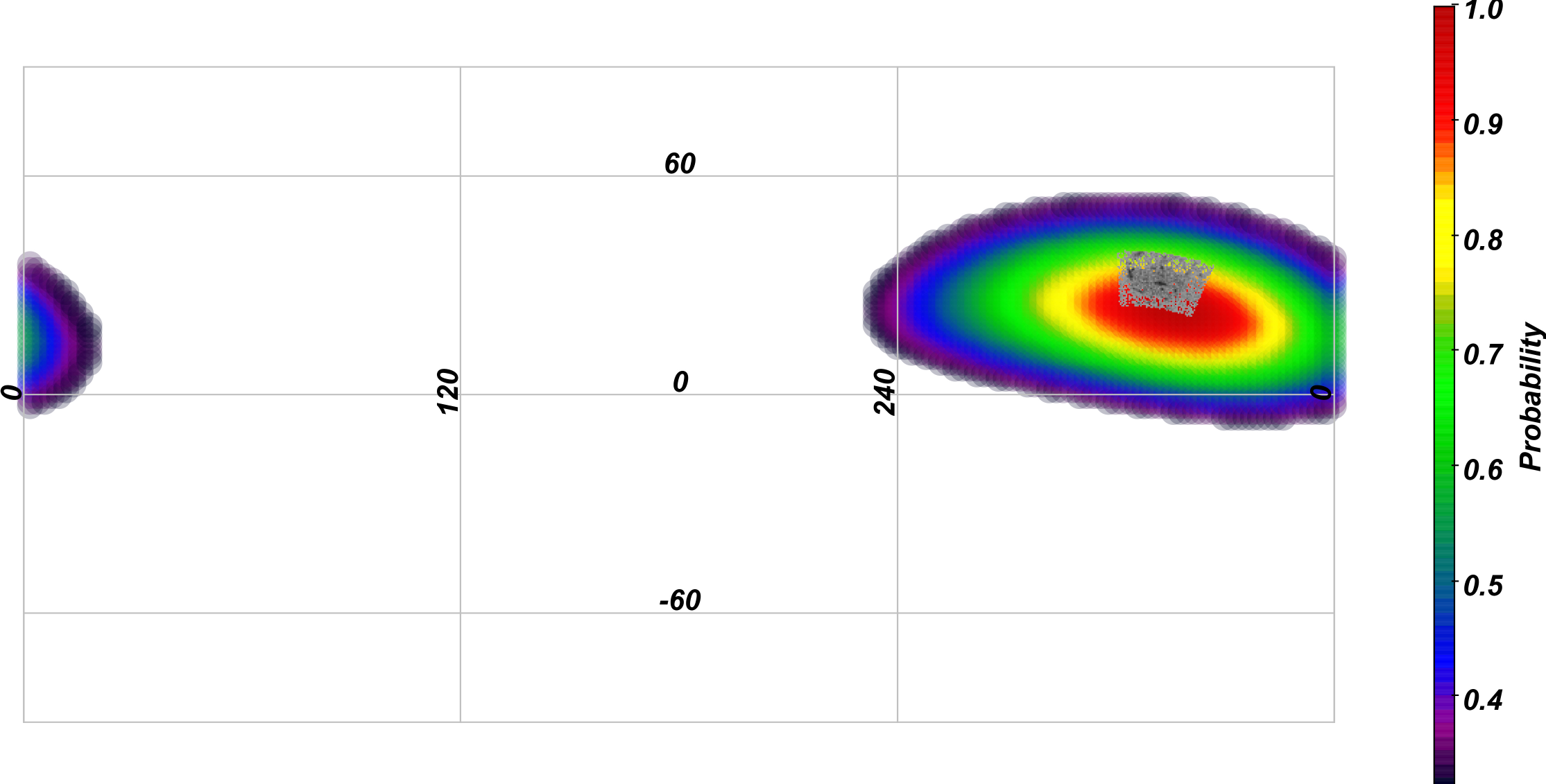}
	
	\caption{\label{fig:i} Probability of bulk flow direction in galactic longitude $l$ and galactic
		latitude $b$ for $0.045<z<0.055$. The most probable direction pointing towards  $(l,b)=(312^{+14}_{-18},26^{+12}_{-8})$. The gray circle corresponds to the spatial extent of the Shapley super cluster supercluster(SSC), providing contextual information for the observed bulk flow patterns. The SSC position is in the direction of bulk flow at (1-$\sigma$) in this redshift. 
	}
\end{figure}

Clusters of galaxies at lower redshifts are easily discerned, notably the Shapley supercluster at redshifts below 0.06. However, as we extend beyond \(z>0.06\), the spatial distribution undergoes a flattening phenomenon attributed to uncertainties, with photometric redshifts gradually assuming dominance. The incorporation of optical or mid-infrared photometry from comprehensive all-sky surveys, exemplified by (WISE), holds the potential to substantially enhance photometric redshift accuracy, thereby enabling the meticulous reconstruction of the local universe beyond the 300–400 Mpc threshold. Moreover, large redshift surveys, such as the notable 6dFGS, furnish precise distance estimates across expansive celestial domains, intricately refining our perspective on the cosmic web. For resdshift $0.4<z<0.6$, there is a most massive supercluster.(\cite{Shimakawa}) The King Ghidorah Supercluster (KGSc) stands as the most massive galaxy supercluster ever discovered, situated approximately 1.3 billion light-years away from Earth. Comprising at least 15 massive galaxy clusters and interconnected filaments, this cosmic colossus is a testament to the grandeur of the universe.
The sheer magnitude of the KGSc is awe-inspiring, boasting a mass of $10^{16}$ solar masses. This extraordinary mass is tenfold greater than our local supercluster, the Laniakea Supercluster. The KGSc spans a colossal distance of about 400 megaparsecs (1.3 billion light-years), showcasing the vastness of its cosmic influence. The groundbreaking discovery of the KGSc in 2022, utilizing data from the Subaru Telescope's Hyper Suprime-Cam (HSC) survey, has significantly expanded our understanding of cosmic structures. The HSC survey, with its ability to detect galaxies over immense distances, unveiled the intricacies of the KGSc, shedding light on its composition and dimensions.
This revelation prompts profound questions about the universe's formation and evolution. The KGSc's unprecedented scale challenges existing cosmological paradigms, hinting at the possibility of undiscovered, even more massive superclusters awaiting exploration. In essence, the KGSc serves as a celestial beacon, guiding astronomers to explore the depths of the cosmos and unravel its mysteries. Figure 12 reveals the direction of KGSc and direction of bulk flow obtained wiyh chameleon model.
\begin{figure}[tbp]
	\centering 
	\includegraphics[width=0.7\textwidth,height=.35\textheight]{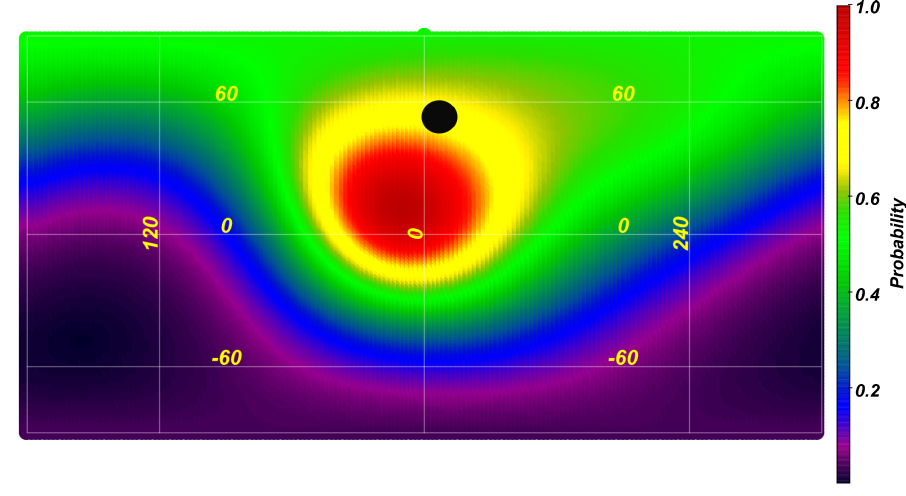}
	
	\caption{\label{fig:i}Probability of bulk flow direction in galactic longitude $l$ and galactic
		latitude $b$ for $0.4<z<6$. The most probable direction pointing towards $(l,b)=6^{+11}_{-10},+15^{+21}_{-19})$ which is broad agreement with \cite{Yarahmadi}. The black circle denotes the direction of The King Ghidorah supercluster. The KGS position is in direction $(l,b)=355,+55)$ in galactic coordinates.}
\end{figure}
	
After conducting an analysis of the data obtained at redshift $0.1<z<1.4$, we have discovered that the bulk flow within this redshift is aligned with the direction of the dark energy dipole. This observation suggests that there may be a correlation between the two phenomena, and further research is warranted to explore this possibility. The implications of such a correlation could have significant implications for our understanding of the nature of dark energy, as well as its effects on the distribution of matter within the universe.

\begin{figure}[tbp]
	\centering 
	\includegraphics[width=0.7\textwidth,height=.35\textheight]{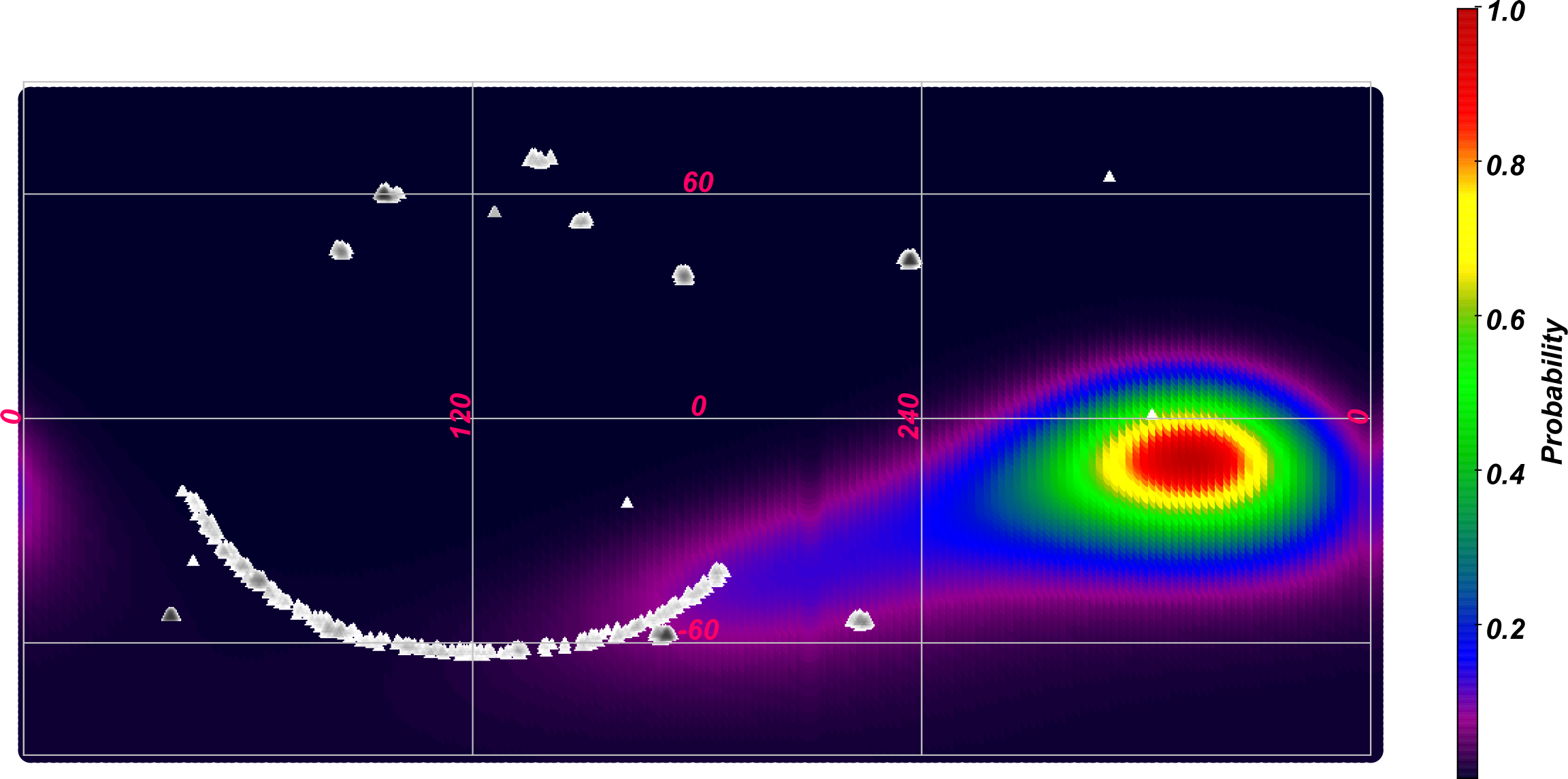}
	
	\caption{\label{fig:i}This figure visualizes the probability distribution of bulk flow direction within the Milky Way, mapped in galactic longitude $l$ and latitude $l$ for redshifts $0.1<z<1.4$. The most probable direction of bulk flow aligns with $(l,b)=(306^{+13}_{-12},-12^{+11}_{-9})$. White triangles represent the distribution of supernovae, which contribute to the data used for this analysis.}
\end{figure}

\section{Comparison of bulk flow direction with dark energy dipole direction} \label{subsubsec:hide}
 Since one of the most likely causes of the universe's accelerated expansion is dark energy, studies  have suggested that dark energy is distributed as a dipole, which causes the anisotropic expansion of the universe and creates a preferred direction for cosmology(\cite{Antoniou};\cite{Yang}).
On the other hand, it seems that the direction of bulk flow on a large scale is close to the  dark energy dipole. As such to investigate this issue, we calculate the  dark energy dipole in the Chameleon model. Numerous studies have investigated the  dark energy dipole in a variety ways (\cite{Antoniou};\cite{Yang}) here using equation (2-1) in reference (\cite{Antoniou}) and equation (5-22) in reference (\cite{Salehi1}). We deal with dipoles (details and equations of dark energy in redshift less than 0.1 equivalent to approximately $300h^{-1}Mpc$ distance) are given in the references mentioned. In Figure 14, the probability distribution function in region $(1-\sigma)$ for the dipole direction of dark energy  in galactic coordinates is plotted and its direction is compared with the direction of the bulk flow in the same region. As can be seen from this figure, these two regions have a good overlap, indicating a unifying convergence between the bulk flow direction and the direction of dark energy dipole at distances greater than $150h^{-1}Mpc$. With compare the figure (1 to 6); it can therefore be concluded that the bulk flow converges in the lower redshift and at distances less than ($150h^{-1}Mpc$) local cluster motion. However as the distance increases, the degree of convergence decreases and the direction of the bulk flow close to the dark energy dipole.
\begin{figure}[!]
	\centering 
	\includegraphics[width=.7\textwidth,height=.35\textheight]{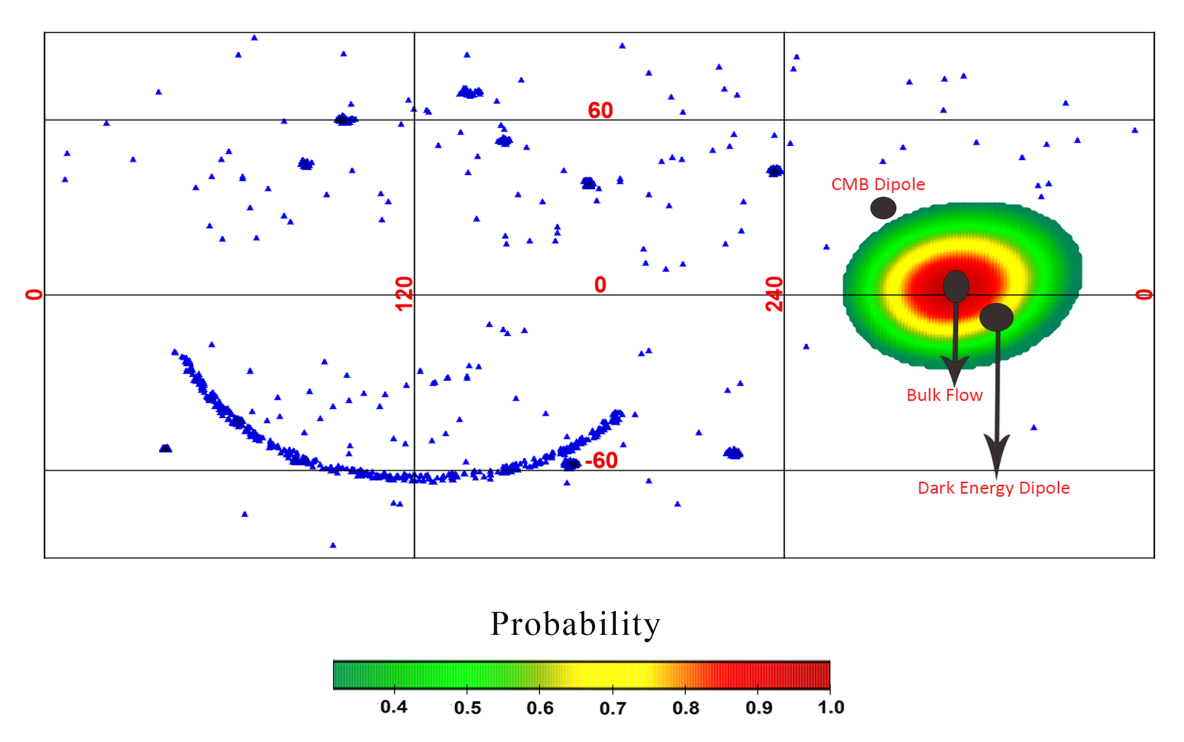}\ \
	\caption{\label{fig:i}Comparison of bulk flow direction and  direction of dark energy dipole in region $(1-\sigma)$. As we can see in this figure, the direction of dark energy diploe is in consist with the direction of bulk flow at $(1-\sigma)$. The blue triangles denotes as Supernovas in redshift $0.1<z<1.4$ }
\end{figure}

\section{Conclusion}
It is challenging to ascertain that recent experiments have comprehensively covered all of the chameleon parameter space. Ideally, a combination of new techniques and searches is required to entirely negate the possibility that screened scalars exist in our universe. 
 In the theoretical approach, the Chameleon model yields excellent results and could potentially explain many cosmological phenomena, as well as other scalar fields. The results of our study indicate that the $\beta$ value falls within the expected range, with a magnitude order of unity. This outcome can be attributed to the fact that scalar fields acquire mass within this range of $\beta$, which is contingent upon the local matter density. This topic has been extensively discussed in various publications, including the origin work of the Chameleon model. All of work in below consider the value of $\beta$  in order of unity(\cite{Burrage}; \cite{Khoury} ;\cite{Weltman} ;\cite{Radouane} ; \cite{Upadhye}; \cite{Salehi3}) .

Additionally, the $\alpha$ value obtained from this study is 0.51 which is in broad agreement with the findings presented in the aforementioned publications (\cite{Salehi3})(0.54), and Cosmography of chameleon gravity (\cite{Salehi2}).\color{black}

In this paper, we have shown that Chameleon fields can be the source of the dark energy dipole and CMB dipole. The study of the region $(1-\sigma)$, at redshifts less than 0.035, shows that not only are our results well consistent with studies confirming the bulk flow, but they also confirm the direction of movement of galaxies of the local group in the region $(1-\sigma)$ towards the Hydra-Centaurus supercluster, which is in the same redshift. Similarly, at redshifts less than 0.06, the direction of movement of the local cluster in the region $(1-\sigma)$ aligns with the bulk flow, and these two movements are in turn in the direction of the Shapley supercluster.

The chameleon model justifies the convergence between the direction of the bulk flow and the motion of a local group galaxy with good accuracy $(1-\sigma)$ in the direction of high-density structures such as superclusters. At short distances, it might be the source of all flows, explaining the mass distribution inhomogeneity. Despite this, it is certainly interesting to study the convergence of the direction of the bulk flow with other cosmic anisotropies originating in the early evolution of the universe, potentially responsible for the origin of the bulk flow, the direction of motion of the local group galaxies, and the formation of large structures.

One of these anisotropies observed by the analysis of observational data is dark energy distribution or dark energy dipole anisotropy. Considering that one of the most probable causes of the accelerated expansion of the universe is dark energy, studies confirm that dark energy is distributed as a bipolar field, causing the anisotropic expansion of the universe.

In this paper, the direction of the dark energy dipole in the Chameleon model is obtained in the direction of galactic coordinates $(l, b) = (324^{+18}_{-18}, -4^{+12}_{-12})$, which is in good agreement with the direction of the bulk flow in the region $(1-\sigma)$. We investigate the relation between the chameleon field and the dark energy dipole. The chameleon field changes its value depending on the mass of a region in space. In regions with a high mass of matter, its mass is high, and in regions with a low mass of matter, its mass changes and becomes lower. Thus, it can be concluded that the distribution of matter in the universe is heterogeneous.

The chameleon field can explain the accelerated expansion of the universe with modifications in gravity. The chameleon particles can interact with matter. Thus, considering that matter is not uniformly distributed in the Universe, the chameleon field becomes heterogeneous. It seems that we can find the possibility of a relationship between the chameleon field and the dark energy dipole.

To prove this relation, we used redshift tomography, focusing on $0.016<z<0.03$, $0.045<z<0.055$, $0.4<z<0.6$, $0.1<z<1.4$ redshift shells for calculating the bulk flow direction. As mentioned above, the density of the chameleon field in regions with high matter density increases, causing non-uniform gravity, which can create a preferred direction in the universe.

 Several significant galactic superclusters, such as Perseus-Pisces (PP), Shapley, The king Ghidorah, are in these redshifts. 
 In  redshift $0.016<z<0.03$, we indicated that the direction of the PP supercluster in galactic coordinates is presented alongside the direction of the bulk flow obtained from the chameleon model. Our numerical simulations reveal that the bulk flow exhibits a direction characterized by galactic longitude ($l$) and galactic latitude ($b$) values of $(l, b) = (112^\circ \pm 12^\circ, -15^\circ \pm 13^\circ)$. As depicted in figure 9, there is a good agreement between the direction of the bulk flow and the position of the PP supercluster. In  redshift $0.045<z<0.055$,  the most probable direction pointing towards  $(l,b)=(312^{+14}_{-18},26^{+12}_{-8})$. The SSC position is in the direction of bulk flow at (1-$\sigma$) in this redshift. For resdshift $0.4<z<0.6$, there is a most massive supercluster (\cite{Shimakawa}). The King Ghidorah Supercluster (KGSc) stands as the most massive galaxy supercluster ever discovered, situated approximately 1.3 billion light-years away from Earth. The most probable direction of bulk flow towards $(l,b)=6^{+11}_{-10},+15^{+21}_{-19})$ which is broad agreement with \cite{Yarahmadi}. The KGS position is in direction $(l,b)=355,+55)$ in galactic coordinates. This massive super cluster is in the bulk flow direction in this redshift.
 \color{black}

  Our results demonstrate that the direction of bulk flow in each redshift is in broad agreement with the position of the related supercluster. These studies reveal that the Chameleon model can be a good option to justify the anisotropy on both small and large scales. Given the properties of these fields, they may also be a good justification for spatial anisotropies of the fine structure constant (fine structure constant dipole), which we intend to investigate in future studies.


\bibliography{sample631}{}
\bibliographystyle{aasjournal}



\end{document}